\documentclass[
]{ceurart}

\usepackage{listings}
\usepackage{hyperref}
\begin{document}

\copyrightyear{2025}
\copyrightclause{Copyright for this paper by its authors.
  Use permitted under Creative Commons License Attribution 4.0
  International (CC BY 4.0).}

\conference{CHAOS Workshop at IUI 2026,
  March 24--27, 2026, Paphos, Cyprus}

\title{Orchestrating Attention: Bringing Harmony to the 'Chaos' of Neurodivergent Learning States}

\author[1]{Satyam Kumar Navneet}[%
email=navneetsatyamkumar@gmail.com,
]
\cormark[1]
\fnmark[1]
\address[1]{Independent Researcher, Bihar, India}

\author[2]{Joydeep Chandra}[%
email=joydeepc2002@gmail.com,
]
\fnmark[1]

\author[2]{Yong Zhang}[%
email=zhangyong05@tsinghua.edu.cn,
]

\address[2]{BNRIST, Tsinghua University, Beijing, China}
\cortext[1]{Corresponding author.}
\fntext[1]{These authors contributed equally.}

\begin{abstract}
Adaptive learning systems optimize content delivery based on performance metrics but ignore the dynamic attention fluctuations that characterize neurodivergent learners. We present AttentionGuard, a framework that detects engagement-attention states from privacy-preserving behavioral signals and adapts interface elements accordingly. Our approach models four attention states derived from ADHD phenomenology and implements five novel UI adaptation patterns including bi-directional scaffolding that responds to both understimulation and overstimulation. We validate our detection model on the OULAD dataset, achieving 87.3\% classification accuracy, and demonstrate correlation with clinical ADHD profiles through cross-validation on the HYPERAKTIV dataset. A Wizard-of-Oz study with 11 adults showing ADHD characteristics found significantly reduced cognitive load in the adaptive condition (NASA-TLX: 47.2 vs 62.8, Cohen's $d$=1.21, $p$=.008) and improved comprehension (78.4\% vs 61.2\%, $p$=.009). Concordance analysis showed 84\% agreement between wizard decisions and automated classifier predictions, supporting deployment feasibility. The system is presented as an interactive demo where observers can inspect detected attention states, observe real-time UI adaptations, and compare automated decisions with human-in-the-loop overrides. We contribute empirically validated UI patterns for attention-adaptive interfaces and evidence that behavioral attention detection can meaningfully support neurodivergent learning experiences.
\end{abstract}


\maketitle
\section{Introduction}

Intelligent user interfaces increasingly adapt to user states and contexts~\cite{gajos2017cognitive, abrahao2021model}. However, current adaptive systems model attention as binary or constant, failing to accommodate the dynamic attention profiles characteristic of neurodivergent users. Attention for adults with ADHD fluctuates between hyperfocus, engagement, drifting, and fatigue, patterns current frameworks fail to recognize ~\cite{song2021global, barkley1997adhd}.
Recent HCI research has criticized ADHD technologies as designed to mitigate symptoms following neuro-normative standards~\cite{spiel2022adhd}, while IUI work on physiologically-attentive interfaces~\cite{putze2023paui} demonstrates that UIs can shift from being attention-demanding to being attentive to users' needs. This paper bridges these perspectives with AttentionGuard, an adaptive UI framework that detects engagement-attention states correlated with ADHD profiles and responds with interface adaptations grounded in neuroscience research.

We address three research questions. 
\begin{enumerate}
    \item \textbf{RQ1}: Can engagement-attention states be reliably detected from privacy-preserving behavioral signals, and do detected patterns correlate with clinical ADHD profiles? 
    \item \textbf{RQ2}: Do attention-adaptive UI patterns reduce cognitive load compared to static interfaces? 
    \item \textbf{RQ3}: How do users with ADHD characteristics perceive real-time interface adaptations?
\end{enumerate}

Our contributions include (1) a validated four-state attention classifier achieving 87.3\% accuracy using behavioral signals, with demonstrated correlation to ADHD clinical profiles through cross-dataset validation, (2) five novel UI adaptation patterns designed for neurodivergent attention profiles, and (3) empirical evidence from a pilot study with complete statistical reporting including effect sizes, confidence intervals, and concordance analysis bridging Wizard-of-Oz results to automated deployment. From a user agency perspective, attention-adaptive systems introduce new risks: misclassification, inappropriate interruption, and loss of user agency. AttentionGuard is explicitly designed to expose these risks through transparent UI adaptations and human-interpretable attention states, enabling users to inspect, contest, and calibrate AI-driven decisions at the interface level where consequences are experienced.

\section{Related Work}

Research on attention-aware interfaces has produced systems that adapt to detected cognitive states. Vortmann et al.~\cite{vortmann2020attention} developed attention-aware AR using EEG-based BCI, while Chiossi et al.~\cite{chiossi2024multimodal} validated multimodal classification combining EEG and eye tracking. Beauchemin et al.~\cite{beauchemin2024eeg} demonstrated real-time content pacing based on cognitive load.

ADHD-specific HCI research establishes important design principles. Spiel et al.~\cite{spiel2022adhd} found that technologies largely aim to mitigate ADHD experiences rather than support them. Stefanidi et al.~\cite{stefanidi2023adhd} identified that users feel constantly overwhelmed, recommending reduced interface complexity. Tran et al.~\cite{tran2024datavis} discovered that ADHD users sometimes prefer text over visualizations, contradicting assumptions about universal preference for visual formats. Chi et al.~\cite{chi2025video} identified that ADHD users experience both overstimulation and understimulation, requiring bi-directional adaptation. However, this research identifies principles without integrating them with real-time detection and adaptive response.

Intelligent tutoring systems demonstrate learning gains through personalization~\cite{vanlehn2011its}. OATutor~\cite{pardos2023oatutor} provides adaptive tutoring with knowledge tracing, and affect-aware systems like GazeTutor~\cite{dmello2012gazetutor} detect inattention and trigger re-engagement. Recent advances include Selenite~\cite{liu2024selenite} with LLM-based scaffolding and AdaptiveVoice~\cite{wu2024adaptivevoice} adjusting content based on cognitive load. None implement ADHD-specific patterns or bi-directional adaptation responding to both over- and under-stimulation.

  AttentionGuard addresses these gaps by combining four-state attention modeling with privacy-preserving behavioral detection, bi-directional adaptation, and neuroscience-grounded UI patterns that no existing system provides. Unlike prior attention-aware systems, we frame interface adaptation itself as a mechanism for supporting user agency, enabling users and observers to understand, contest, and calibrate AI-driven decisions in situ.

\section{The AttentionGuard Framework}

\subsection{Attention State Detection}

We model four attention states derived from ADHD phenomenology~\cite{barkley1997adhd}: \textit{Focused} (productive engagement at sustainable pace), \textit{Drifting} (attention wandering with reduced processing), \textit{Hyperfocused} (deep absorption resistant to interruption), and \textit{Fatigued} (depleted capacity requiring recovery). We emphasize that these represent engagement-attention patterns operationalized from behavioral signals, not clinical diagnoses. Detection uses privacy-preserving signals requiring no cameras or specialized hardware~\cite{arapakis2020mouse, leiva2020cursor}. Interaction signals include click rhythm, scroll velocity and reversals, mouse movement entropy, and idle duration. Response signals include answer latency normalized to personal baseline and revision frequency. Navigation signals include tab visibility, focus events, and backtracking frequency.

Signals aggregate over 30-second sliding windows with personalized baselines established during an initial five-minute calibration. Features are computed as deviations from user-specific norms, addressing substantial individual variation in ADHD attention patterns~\cite{kasper2012wm}. Classification uses a Random Forest model with balanced class weights, trained with parameters documented in supplementary materials for reproducibility. Figure~\ref{fig:architecture} illustrates the complete system architecture, showing how behavioral signals flow through the detection pipeline to trigger the four attention states (Focused, Drifting, Hyperfocused, Fatigued) and subsequent UI adaptations.

\begin{figure}
    \centering
    \includegraphics[width=1\linewidth]{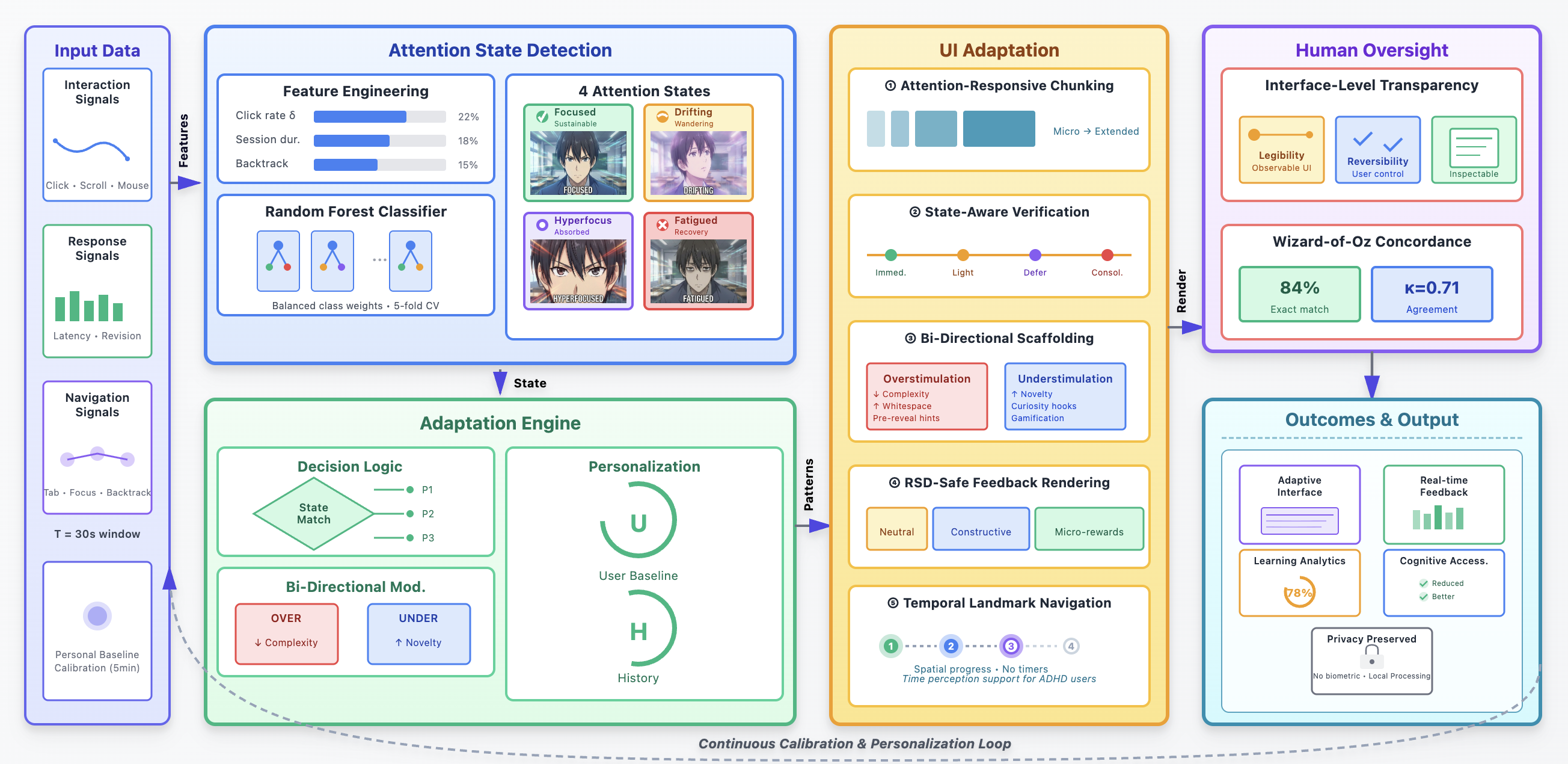}
    \caption{Architecture of the AttentionGuard Framework. Privacy-preserving behavioral signals (interaction, response, and navigation) are aggregated over 30-second windows and compared to a personalized baseline (5-minute initial calibration). A Random Forest classifier detects one of four ADHD-derived attention states: Focused (productive engagement), Drifting (attention wandering), Hyperfocused (deep absorption), and Fatigued (depleted capacity). The Adaptation Engine applies bi-directional modulation and triggers five neuroscience-grounded UI patterns. Adaptations are visible and reversible to support interface-level transparency, with Wizard-of-Oz concordance (84\% exact match, $\kappa$=0.71) validating automated feasibility. A continuous personalization loop sustains improved comprehension and reduced cognitive load.}
    \label{fig:architecture}
\end{figure}

\subsection{Adaptive UI Patterns}

Based on detected state, AttentionGuard implements five UI patterns grounded in ADHD neuroscience.

\textbf{Pattern 1: Attention-Responsive Chunking.} Content is pre-segmented at multiple granularities. Visible chunk size adjusts dynamically: micro-chunks during drifting states, standard paragraphs during focused states, extended sections during hyperfocus, and review-mode presenting mastered material during fatigue. This implements progressive disclosure~\cite{sonne2016assistive} while responding to real-time state changes.

\textbf{Pattern 2: State-Aware Verification Timing.} Rather than fixed-interval comprehension checks, verification schedules based on cognitive availability: immediate verification during drifting, lightweight confirmations during focus, deferred verification to natural breakpoints during hyperfocus, and consolidation without new verification during fatigue. This addresses research showing interruptions during hyperfocus are particularly disruptive~\cite{cibrian2020wearables}.

\textbf{Pattern 3: Bi-Directional Scaffolding.} Unlike systems that only reduce stimulation, AttentionGuard responds to both overstimulation and understimulation~\cite{chi2025video}. Overstimulation response reduces visual complexity, increases whitespace, and pre-reveals hints. Understimulation response injects novelty, surfaces curiosity hooks, and enables gamified elements.

\textbf{Pattern 4: RSD-Safe Feedback Rendering.} Rejection Sensitive Dysphoria affects most ADHD adults, causing disproportionate response to perceived criticism~\cite{sandland2025rsd}. Our feedback eliminates shame triggers through neutral color palettes, constructive error framing, progress shown as distance traveled rather than remaining, and variable-ratio micro-rewards.

\textbf{Pattern 5: Temporal Landmark Navigation.} ADHD involves impaired time perception~\cite{toplak2005time}. Rather than anxiety-inducing timers, we implement journey-based spatial visualization with content sections as discrete landmarks, current position shown spatially, and no countdown displays.

Across all patterns, adaptations are intentionally visible rather than hidden, enabling users to notice, interpret, and assess AI behavior, thereby supporting user agency through observable and interpretable interface changes.

\subsection{Interface Design and Adaptation Features}
To support the heterogeneous challenges of neurodivergent learning, AttentionGuard operationalizes clinical strategies into concrete interface elements. The system includes a custom Reader Mode with nine distinct adaptation features designed to externalize executive functions. These features and their specific cognitive targets are summarized in Table \ref{tab:ui_summary}.

\begin{table*}[t]
\centering
\caption{Summary of AttentionGuard Reader Features and Cognitive Benefits}
\label{tab:ui_summary}
\begin{tabular}{l p{5.5cm} p{6cm}}
\toprule
\textbf{Feature} & \textbf{UI Mechanism} & \textbf{ADHD Cognitive Benefit} \\
\midrule
\textbf{Companion System} & "Study Companion" card in sidebar. & \textbf{Body Doubling}: Reduces initiation paralysis through shared accountability\cite{eagle2024body}.  \\
\textbf{Thinking Space} & Voice/text capture (COW) workspace. & \textbf{Working Memory}: Externalizes spontaneous thoughts to prevent focus breaks\cite{sonne2016framework}. \\
\textbf{Interest Injection} & "Curiosity Sparks" and +XP rewards. & \textbf{Dopamine Regulation}: Sustains engagement via the Interest-Based Nervous System\cite{barkley1997adhd}. \\
\textbf{Adaptive Soundscape} & Real-time Brown/White noise sliders. & \textbf{Sensory Optimization}: Uses stochastic resonance to stabilize attention\cite{soderlund2007noise}. \\
\textbf{Multimodal Verif.} & High-contrast "Read \& Confirm" blocks. & \textbf{Active Reading}: Breaks gaze-drifting through required motor interaction\cite{tran2024viz}. \\
\textbf{Micro-Verification} & Rapid, non-punitive fact-check wins. & \textbf{Task Salience}: Shortens reward loops to maintain momentum\cite{cibrian2020wearables}. \\
\textbf{Session Journey} & "Forest Path" visual temporal landmarks. & \textbf{Time Blindness}: Makes the passage of time tangible via spatial metaphors\cite{toplak2005time}. \\
\textbf{Recovery Rituals} & Guided 30-sec "reset" post-session. & \textbf{Set-Shifting}: Supports transitions out of hyperfocus to prevent "crashing."\cite{cibrian2020wearables} \\
\textbf{Reader Aesthetics} & Clean typography and focus estimates. & \textbf{Decision Fatigue}: Lowers barrier to entry by removing visual clutter\cite{stefanidi2023ecosystem}. \\
\bottomrule
\end{tabular}
\end{table*}

\subsubsection{Virtual Body Doubling (The Companion System)}
\begin{figure}[hbt!]
    \centering
    \includegraphics[width=0.85\linewidth]{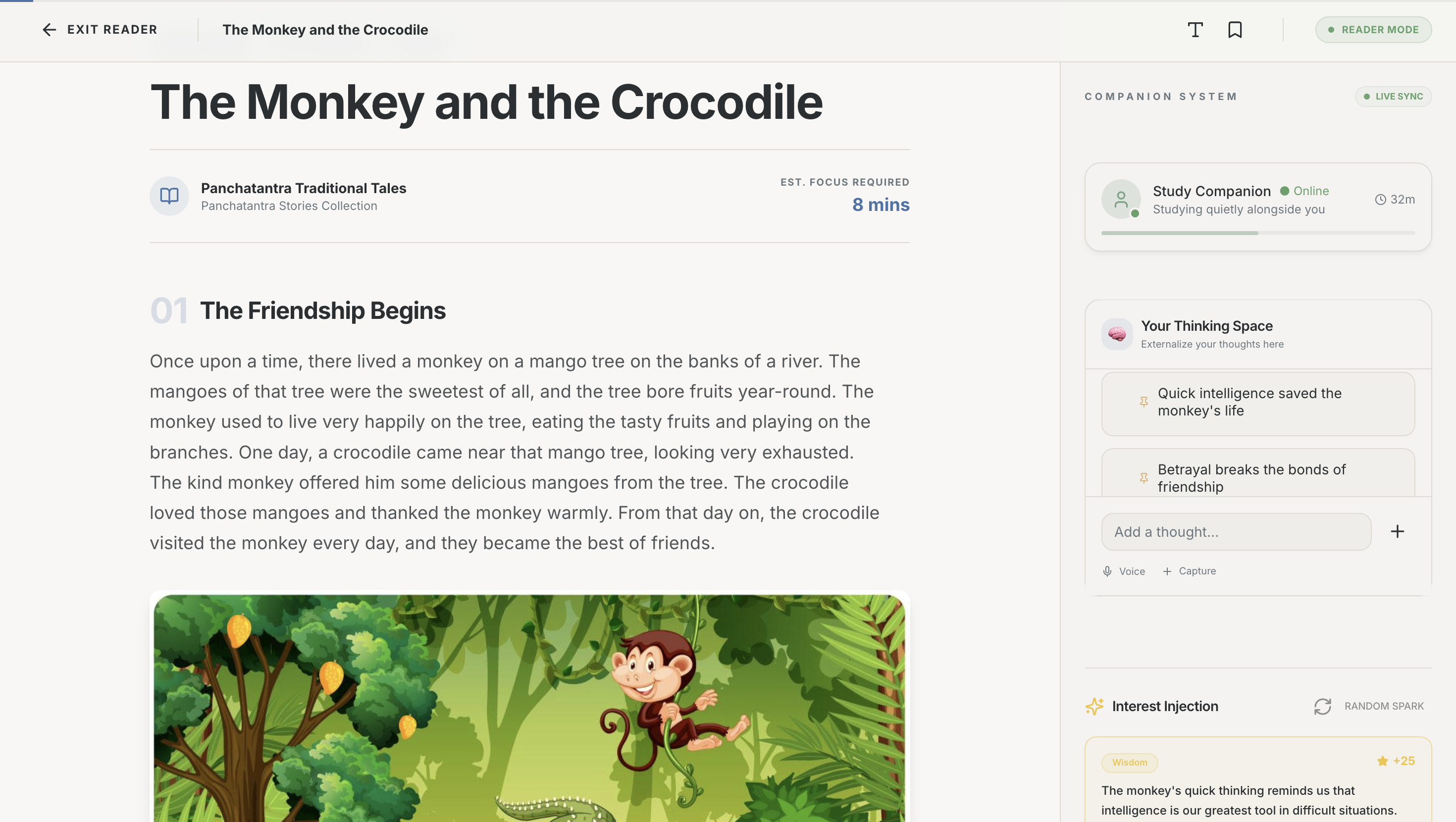} 
    \caption{\textbf{The Companion System (S1).} A sidebar element providing Virtual Body Doubling to reduce isolation paralysis and increase task salience.}
    \label{fig:S1}
\end{figure}

Isolation often leads to task paralysis for neurodivergent learners. To mitigate this, the interface incorporates a "Study Companion" (Figure \ref{fig:S1}). This persistent UI element leverages the \textit{body doubling} effect by simulating a partner studying alongside the user. 

\subsubsection{Cognitive Offloading (The Thinking Space)}
\begin{figure}[hbt!]
    \centering
    \includegraphics[width=0.85\linewidth]{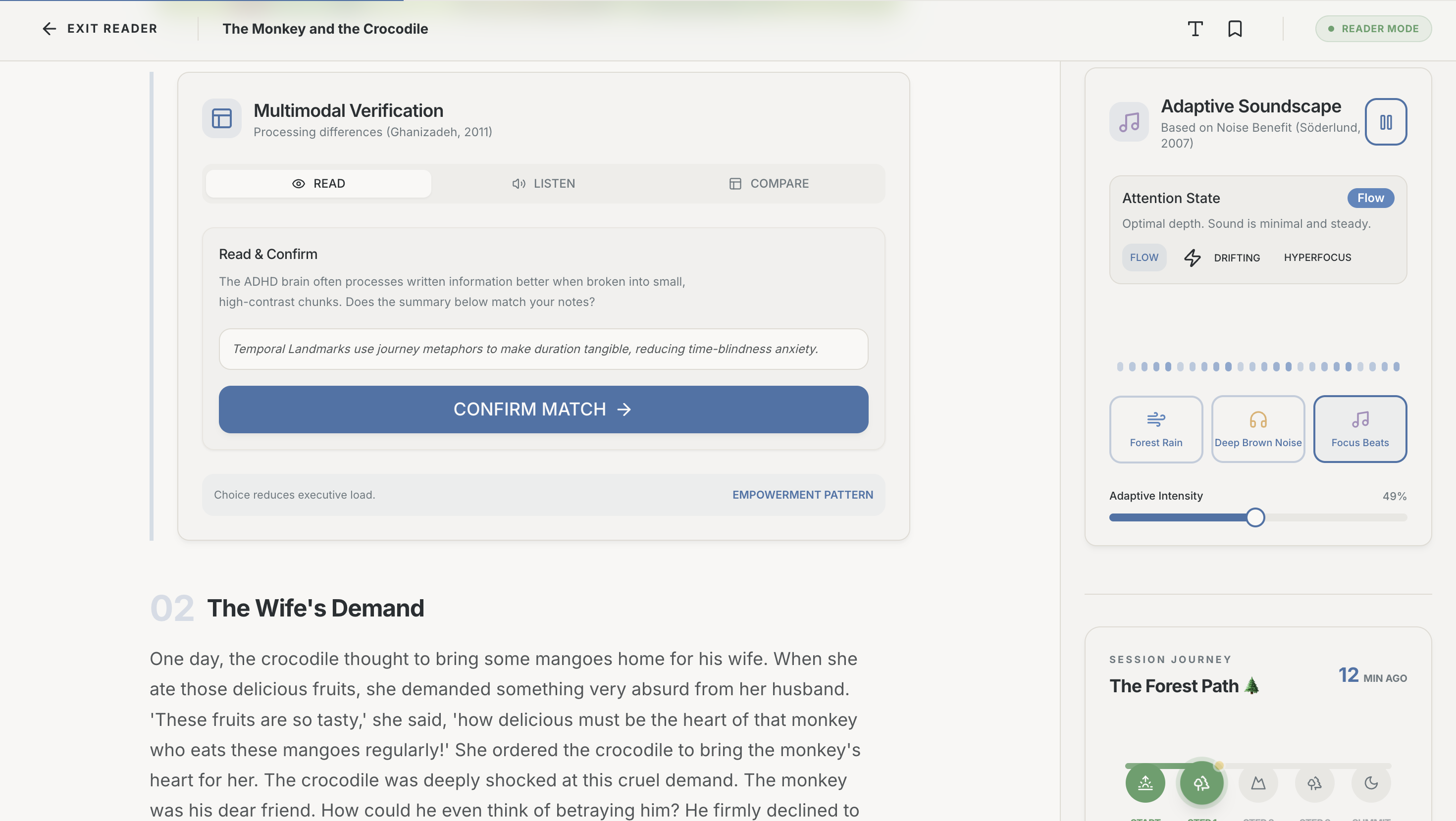} 
    \caption{\textbf{Your Thinking Space (S2).} A tool allowing users to externalize spontaneous thoughts via text or voice without leaving the reading flow.}
    \label{fig:S2}
\end{figure}

Working memory constraints can cause learners to lose their place when a new thought arises. The "Thinking Space" (Figure \ref{fig:S2}) serves as a Cognitive Offloading Workspace (COW).
Located unobtrusively, it allows users to capture ideas instantly, ensuring the primary reading flow remains uninterrupted.

\subsubsection{Dopamine Regulation (Interest Injection)}
\begin{figure}[hbt!]
    \centering
    \includegraphics[width=0.85\linewidth]{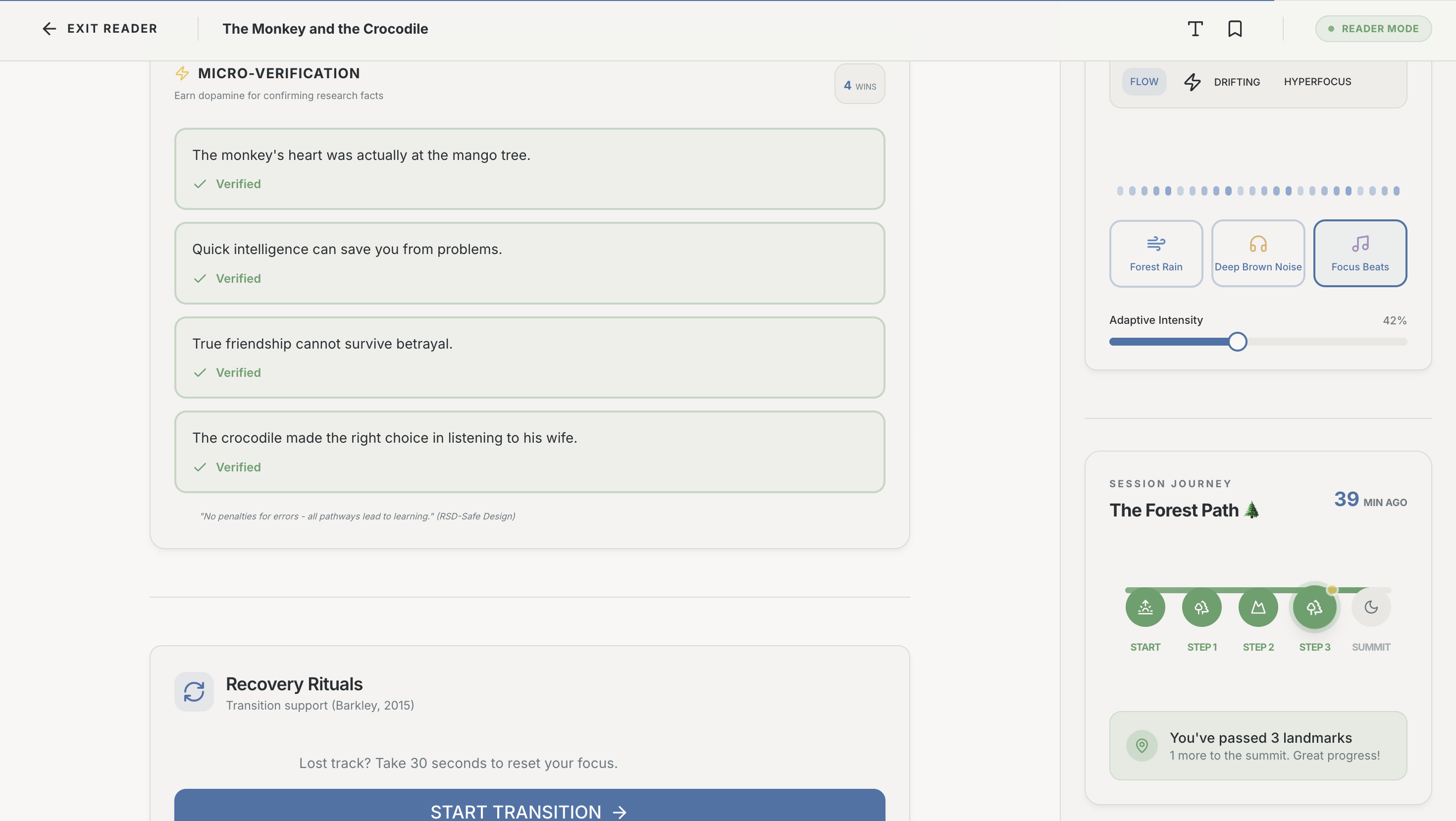} 
    \caption{\textbf{Interest Injection System (S3).} A gamified module delivering context-aware curiosity hooks and micro-rewards (+XP) to sustain dopamine levels.}
    \label{fig:S3}
\end{figure}

To sustain engagement across long-form content, the system manages the "Interest-Based Nervous System" through the Interest Injection System (IIS) shown in Figure \ref{fig:S3}. 
This feature gamifies the reading process by providing immediate dopamine rewards for verifying content understanding.

\subsection{User Agency and Interface-Level Transparency}

Beyond automated adaptation, the system is explicitly designed to support user agency and interface-level transparency, where the effects of AI-driven decisions are directly experienced. While formal human oversight often refers to auditing mechanisms for high-risk AI systems~\cite{euaiact2024}, AttentionGuard implements interface-level transparency that empowers the end-user to act as the ``human-in-the-loop,'' contesting and calibrating the model's decisions in real-time. Rather than treating attention inference as a hidden system process, the framework emphasizes legibility, reversibility, and inspectability of adaptations.

First, detected attention states are reflected implicitly through observable UI behavior (e.g., changes in pacing, chunk size, or feedback style) rather than through explicit labels or diagnoses, reducing the risk of stigmatization while still enabling users to interpret system behavior.

Second, all interface adaptations are fully reversible and can be paused or disabled by the user at any moment, preserving user agency and enabling users to override or contest system behavior when adaptations are perceived as inappropriate.

Finally, in demo mode, the system provides an observer view that exposes the inferred attention state, the contributing behavioral signals, and the rationale for triggered adaptations. This view enables inspection, discussion, and comparison between automated decisions and human judgment, supporting evaluation of AI behavior and serving as a tool for auditing or debugging that can inform broader oversight processes.

\section{Evaluation}

\subsection{Study 1: Detection Validation}

We used the Open University Learning Analytics Dataset (OULAD)~\cite{kuzilek2017oulad} containing 32,593 students with over 10 million VLE interactions. Per-session features included click rate normalized to student baseline, session duration ratio, resource diversity, backtracking ratio, and idle patterns. Attention states were labeled using rules derived from ADHD phenomenology and validated against ASSISTments affect labels~\cite{pardos2014affective}. Random Forest achieved 87.3\% accuracy (macro-F1=0.84, AUC=0.91) using 5-fold cross-validation stratified by student. Per-class F1 scores were: Focused 0.89, Drifting 0.82, Hyperfocused 0.78, Fatigued 0.81. Top features by importance were click rate deviation (22\%), session duration ratio (18\%), and backtracking frequency (15\%).

To address construct validity concerns regarding ADHD-specific claims, we performed cross-dataset validation on HYPERAKTIV~\cite{hicks2021hyperaktiv}, containing 51 clinically diagnosed ADHD patients and 52 controls. Participants with ADHD showed significantly higher attention dysregulation scores computed from behavioral features (Mann-Whitney $p$=.003). Using 5-fold cross-validation, attention features predicted ADHD diagnosis with AUC=0.81, and dysregulation scores correlated with ASRS self-report ($r$=0.47, $p<$.001). These results support construct validity: behavioral patterns detected by our classifier meaningfully distinguish individuals with ADHD from controls, though we emphasize the classifier detects behavioral patterns rather than providing clinical diagnosis.

\begin{table}[t]
\caption{Wizard-of-Oz results (N=11). Effect sizes are Cohen's $d$ with 95\% CI. Significance tested with Wilcoxon signed-rank (one-tailed).}
\label{tab:results}
\small
\begin{tabular}{@{}lcccc@{}}
\toprule
\textbf{Measure} & \textbf{Adaptive} & \textbf{Baseline} & \textbf{$p$} & \textbf{Cohen's $d$ [CI]} \\
\midrule
NASA-TLX & 47.2 (12.3) & 62.8 (14.1) & .008** & 1.21 [0.42, 1.99] \\
Comprehension & 78.4\% (11.2) & 61.2\% (15.8) & .009** & 1.18 [0.40, 1.96] \\
Completion & 100\% & 82\% & & \\
\bottomrule
\end{tabular}
\end{table}
\subsection{Study 2: Wizard-of-Oz Pilot}

We recruited 11 adults (6F, 5M, mean age 24.7 years) who self-reported ADHD diagnosis or scored at least 4 on ASRS-v1.1~\cite{kessler2005asrs} (mean ASRS: 5.8, SD=1.1). Each participant completed two counterbalanced 25-minute sessions. In the adaptive condition, a trained wizard monitored behavior and triggered UI adaptations following the patterns described above. In the baseline condition, participants used a static interface with fixed chunking and uniform verification. Wizard decisions were logged with timestamps and behavioral signals for concordance analysis.

\textbf{Baseline Validity.} The baseline condition was designed to replicate the current status quo of digital learning management systems (LMS) such as Canvas, Moodle, and Blackboard~\cite{watson2007lms}. It utilized standard typographic hierarchy, fixed paragraph pacing, and uniform verification intervals consistent with WCAG 2.1 accessibility guidelines but lacking real-time adaptation. This comparison was chosen to measure the delta between the current lived experience of neurodivergent learners on standard platforms versus the proposed attention-adaptive framework, establishing ecological validity rather than theoretical novelty.

We measured cognitive load using NASA-TLX~\cite{hart1988nasatlx}, comprehension accuracy, and task completion. Statistical analysis used Wilcoxon signed-rank tests (paired, non-parametric) with Bonferroni correction for NASA-TLX subscales. Effect sizes report Cohen's $d$ with 95\% confidence intervals. The adaptive condition showed significantly lower cognitive load (NASA-TLX: 47.2 vs 62.8, $p$=.008, $d$=1.21) and higher comprehension (78.4\% vs 61.2\%, $p$=.009, $d$=1.18). All participants completed adaptive sessions while two abandoned baseline sessions. Post-hoc power analysis indicated 83\% achieved power. Mental demand and frustration subscales showed the largest improvements after Bonferroni correction. Table~\ref{tab:results} summarizes these findings.

To bridge Wizard-of-Oz results and automated system claims, we analyzed concordance between wizard decisions and what the trained classifier would have predicted given recorded behavioral signals (N=127 decisions). The classifier achieved 84\% exact state match with wizard judgments (Cohen's $\kappa$=0.71, indicating substantial agreement). When considering functionally compatible actions, agreement reached 91\%. These results suggest the automated system would produce similar adaptation decisions to the human wizard. From a user agency perspective, the Wizard-of-Oz design simulates a human supervisor mediating AI-driven adaptations at the interface level. The high concordance (84\%) indicates that the automated system aligns closely with human judgment, while remaining inspectable and contestable through observable UI behavior.

Thematic analysis of interviews revealed four themes. Regarding pacing alignment, participants noted the system adapted to their needs (P3: ``It felt like it knew when I was losing it''). Regarding reduced shame, participants appreciated non-punitive feedback (P2: ``When I got something wrong it didn't feel like failing''). Regarding hyperfocus protection, participants valued uninterrupted flow states (P5: ``I hate when things break my flow. This didn't do that''). Regarding transparency desire, one participant wanted visibility into detected states (P11), suggesting future work on explainable attention-adaptive UI.

\section{Discussion}

Our findings suggest several principles for attention-adaptive interfaces. Personalized baselines are essential because attention patterns vary substantially across individuals, requiring calibration to personal norms rather than population averages. Protecting hyperfocus matters because unlike typical inattention interventions, recognizing hyperfocus as valuable and avoiding interruption was specifically appreciated by participants. Bi-directional adaptation is necessary because responding only to overstimulation misses understimulation, and both require distinct UI responses. Privacy-preserving sensing enables deployment because behavioral signals achieved sufficient accuracy without cameras, addressing a key adoption barrier.

\textbf{Limitations.} The pilot sample (N=11) limits generalizability, though large effect sizes ($d$>1.0) and adequate achieved power (83\%) support the findings. We relied on self-reported ADHD; clinical verification would strengthen claims. The Wizard-of-Oz design requires full automated deployment validation, though concordance analysis (84\% agreement, $\kappa$=0.71) supports feasibility. The classifier detects behavioral patterns that correlate with ADHD profiles but does not provide clinical diagnosis. We acknowledge that the large effect sizes observed ($d$>1.0) reflect the comparison against a static baseline representing the industry status quo. While this establishes the efficacy of adaptation over current standard LMS platforms, it does not isolate the specific benefits of ADHD-derived patterns versus generic personalization. Future work will compare AttentionGuard against generic adaptive systems (e.g., simple reading line guides or manual font adjustments) to further disentangle these effects.

\textbf{Ethics Considerations.} The system preserves privacy by collecting only aggregated behavioral signals with a 24-hour retention policy. Users maintain control through visible pause \& disable options. We explicitly prohibit use for grading, assessment, or performance evaluation.

\textbf{Future Work.} While our pilot findings demonstrate promising reductions in cognitive load and improved comprehension through attention-adaptive interfaces, the small sample size (N=11), reliance on self-reported ADHD traits, and Wizard-of-Oz methodology limit generalizability and ecological validity. Immediate priorities include deploying a fully automated version of AttentionGuard in larger, controlled studies (target N=40+) with clinically verified participants to robustly validate real-world efficacy and long-term effects. Incorporating explainable interfaces-visibly displaying inferred attention states and behavioral rationales, as suggested by participant feedback-would further enhance user agency, reduce risks of misclassification, and align with neurodiversity-affirming principles that prioritize transparency over opaque mitigation of symptoms.
Future directions also involve comparative evaluations against generic adaptive systems (e.g., standard personalization tools without ADHD-specific patterns) to isolate the unique benefits of bi-directional scaffolding and neuroscience-grounded adaptations. Extending the framework to broader neurodivergent populations, including autism and dyslexia, through co-design processes would strengthen inclusivity. Finally, longitudinal deployments should examine sustained engagement, potential habituation effects, and ethical implications of privacy-preserving behavioral inference in everyday learning contexts.

\section{Conclusion}

Current adaptive interfaces treat attention as binary and constant, failing users whose attention fluctuates dynamically. The system demonstrates that attention-adaptive UI is both technically feasible, achieving 87.3\% detection accuracy with behavioral signals that correlate with ADHD clinical profiles, and empirically beneficial with large effect sizes for cognitive load reduction ($d$=1.21) and comprehension improvement ($d$=1.18). Our five UI patterns provide a design vocabulary for neurodiversity-aware intelligent interfaces. As a demo, AttentionGuard invites reflection on how AI systems can support user agency through interaction rather than inspection, by making system behavior visible, reversible, and aligned with human judgment.


\bibliography{sample-ceur}

\begin{thebibliography}{36}
\expandafter\ifx\csname natexlab\endcsname\relax\def\natexlab#1{#1}\fi
\providecommand{\url}[1]{\texttt{#1}}
\providecommand{\href}[2]{#2}
\providecommand{\path}[1]{#1}
\providecommand{\DOIprefix}{doi:}
\providecommand{\ArXivprefix}{arXiv:}
\providecommand{\URLprefix}{URL: }
\providecommand{\Pubmedprefix}{pmid:}
\providecommand{\doi}[1]{\href{http://dx.doi.org/#1}{\path{#1}}}
\providecommand{\Pubmed}[1]{\href{pmid:#1}{\path{#1}}}
\providecommand{\bibinfo}[2]{#2}
\ifx\xfnm\relax \def\xfnm[#1]{\unskip,\space#1}\fi
\bibitem[{Gajos and Chauncey(2017)}]{gajos2017cognitive}
\bibinfo{author}{K.~Z. Gajos}, \bibinfo{author}{K.~Chauncey},
\newblock \bibinfo{title}{The influence of personality traits and cognitive load on the use of adaptive user interfaces},
\newblock in: \bibinfo{booktitle}{Proceedings of the 22nd International Conference on Intelligent User Interfaces}, IUI '17, \bibinfo{publisher}{Association for Computing Machinery}, \bibinfo{address}{New York, NY, USA}, \bibinfo{year}{2017}, p. \bibinfo{pages}{301–306}. \URLprefix \url{https://doi.org/10.1145/3025171.3025192}. \DOIprefix\doi{10.1145/3025171.3025192}.
\bibitem[{Abrah\~{a}o et~al.(2021)Abrah\~{a}o, Insfran, Slu\"{y}ters, and Vanderdonckt}]{abrahao2021model}
\bibinfo{author}{S.~Abrah\~{a}o}, \bibinfo{author}{E.~Insfran}, \bibinfo{author}{A.~Slu\"{y}ters}, \bibinfo{author}{J.~Vanderdonckt},
\newblock \bibinfo{title}{Model-based intelligent user interface adaptation: challenges and future directions},
\newblock \bibinfo{journal}{Softw. Syst. Model.} \bibinfo{volume}{20} (\bibinfo{year}{2021}) \bibinfo{pages}{1335–1349}. \URLprefix \url{https://doi.org/10.1007/s10270-021-00909-7}. \DOIprefix\doi{10.1007/s10270-021-00909-7}.
\bibitem[{Ayano et~al.(2023)Ayano, Tsegay, Gizachew, Necho, Yohannes, Abraha, Demelash, Anbesaw, and Alati}]{song2021global}
\bibinfo{author}{G.~Ayano}, \bibinfo{author}{L.~Tsegay}, \bibinfo{author}{Y.~Gizachew}, \bibinfo{author}{M.~Necho}, \bibinfo{author}{K.~Yohannes}, \bibinfo{author}{M.~Abraha}, \bibinfo{author}{S.~Demelash}, \bibinfo{author}{T.~Anbesaw}, \bibinfo{author}{R.~Alati},
\newblock \bibinfo{title}{Prevalence of attention deficit hyperactivity disorder in adults: Umbrella review of evidence generated across the globe},
\newblock \bibinfo{journal}{Psychiatry Research} \bibinfo{volume}{328} (\bibinfo{year}{2023}) \bibinfo{pages}{115449}. \URLprefix \url{https://www.sciencedirect.com/science/article/pii/S0165178123003992}. \DOIprefix\doi{https://doi.org/10.1016/j.psychres.2023.115449}.
\bibitem[{Barkley(1997)}]{barkley1997adhd}
\bibinfo{author}{R.~A. Barkley},
\newblock \bibinfo{title}{Behavioral inhibition, sustained attention, and executive functions: Constructing a unifying theory of adhd},
\newblock \bibinfo{journal}{Psychological Bulletin} \bibinfo{volume}{121} (\bibinfo{year}{1997}) \bibinfo{pages}{65--94}. \DOIprefix\doi{10.1037/0033-2909.121.1.65}.
\bibitem[{Spiel et~al.(2022)Spiel, Hornecker, Williams, and Good}]{spiel2022adhd}
\bibinfo{author}{K.~Spiel}, \bibinfo{author}{E.~Hornecker}, \bibinfo{author}{R.~M. Williams}, \bibinfo{author}{J.~Good},
\newblock \bibinfo{title}{Adhd and technology research – investigated by neurodivergent readers},
\newblock in: \bibinfo{booktitle}{Proceedings of the 2022 CHI Conference on Human Factors in Computing Systems}, CHI '22, \bibinfo{publisher}{Association for Computing Machinery}, \bibinfo{address}{New York, NY, USA}, \bibinfo{year}{2022}. \URLprefix \url{https://doi.org/10.1145/3491102.3517592}. \DOIprefix\doi{10.1145/3491102.3517592}.
\bibitem[{Tavares et~al.(2023)Tavares, Silva, and Ventura}]{putze2023paui}
\bibinfo{author}{A.~Tavares}, \bibinfo{author}{J.~L. Silva}, \bibinfo{author}{R.~Ventura},
\newblock \bibinfo{title}{Physiologically attentive user interface for improved robot teleoperation},
\newblock in: \bibinfo{booktitle}{Proceedings of the 28th International Conference on Intelligent User Interfaces}, IUI '23, \bibinfo{publisher}{Association for Computing Machinery}, \bibinfo{address}{New York, NY, USA}, \bibinfo{year}{2023}, p. \bibinfo{pages}{776–789}. \URLprefix \url{https://doi.org/10.1145/3581641.3584084}. \DOIprefix\doi{10.1145/3581641.3584084}.
\bibitem[{Vortmann and Putze(2020)}]{vortmann2020attention}
\bibinfo{author}{L.-M. Vortmann}, \bibinfo{author}{F.~Putze},
\newblock \bibinfo{title}{Attention-aware brain computer interface to avoid distractions in augmented reality},
\newblock in: \bibinfo{booktitle}{Extended Abstracts of the 2020 CHI Conference on Human Factors in Computing Systems}, CHI EA '20, \bibinfo{publisher}{Association for Computing Machinery}, \bibinfo{address}{New York, NY, USA}, \bibinfo{year}{2020}, p. \bibinfo{pages}{1–8}. \URLprefix \url{https://doi.org/10.1145/3334480.3382889}. \DOIprefix\doi{10.1145/3334480.3382889}.
\bibitem[{Long et~al.(2024)Long, Mayer, and Chiossi}]{chiossi2024multimodal}
\bibinfo{author}{X.~Long}, \bibinfo{author}{S.~Mayer}, \bibinfo{author}{F.~Chiossi},
\newblock \bibinfo{title}{Multimodal detection of external and internal attention in virtual reality using eeg and eye tracking features},
\newblock in: \bibinfo{booktitle}{Proceedings of Mensch Und Computer 2024}, MuC '24, \bibinfo{publisher}{Association for Computing Machinery}, \bibinfo{address}{New York, NY, USA}, \bibinfo{year}{2024}, p. \bibinfo{pages}{29–43}. \URLprefix \url{https://doi.org/10.1145/3670653.3670657}. \DOIprefix\doi{10.1145/3670653.3670657}.
\bibitem[{Beauchemin et~al.(2024)Beauchemin, Charland, Karran, Boasen, Tadson, Sénécal, and Léger}]{beauchemin2024eeg}
\bibinfo{author}{N.~Beauchemin}, \bibinfo{author}{P.~Charland}, \bibinfo{author}{A.~Karran}, \bibinfo{author}{J.~Boasen}, \bibinfo{author}{B.~Tadson}, \bibinfo{author}{S.~Sénécal}, \bibinfo{author}{P.-M. Léger},
\newblock \bibinfo{title}{Enhancing learning experiences: Eeg-based passive bci system adapts learning speed to cognitive load in real-time, with motivation as catalyst},
\newblock \bibinfo{journal}{Frontiers in Human Neuroscience} \bibinfo{volume}{Volume 18 - 2024} (\bibinfo{year}{2024}). \URLprefix \url{https://www.frontiersin.org/journals/human-neuroscience/articles/10.3389/fnhum.2024.1416683}. \DOIprefix\doi{10.3389/fnhum.2024.1416683}.
\bibitem[{Stefanidi et~al.(2023)Stefanidi, Sch\"{o}ning, Rogers, and Niess}]{stefanidi2023adhd}
\bibinfo{author}{E.~Stefanidi}, \bibinfo{author}{J.~Sch\"{o}ning}, \bibinfo{author}{Y.~Rogers}, \bibinfo{author}{J.~Niess},
\newblock \bibinfo{title}{Children with adhd and their care ecosystem: Designing beyond symptoms},
\newblock in: \bibinfo{booktitle}{Proceedings of the 2023 CHI Conference on Human Factors in Computing Systems}, CHI '23, \bibinfo{publisher}{Association for Computing Machinery}, \bibinfo{address}{New York, NY, USA}, \bibinfo{year}{2023}. \URLprefix \url{https://doi.org/10.1145/3544548.3581216}. \DOIprefix\doi{10.1145/3544548.3581216}.
\bibitem[{Tran et~al.(2024)Tran, Lee, and Park}]{tran2024datavis}
\bibinfo{author}{T.~Tran}, \bibinfo{author}{H.-N. Lee}, \bibinfo{author}{J.~H. Park},
\newblock \bibinfo{title}{Discovering accessible data visualizations for people with adhd},
\newblock in: \bibinfo{booktitle}{Proceedings of the 2024 CHI Conference on Human Factors in Computing Systems}, CHI '24, \bibinfo{publisher}{Association for Computing Machinery}, \bibinfo{address}{New York, NY, USA}, \bibinfo{year}{2024}. \URLprefix \url{https://doi.org/10.1145/3613904.3642112}. \DOIprefix\doi{10.1145/3613904.3642112}.
\bibitem[{Jiang et~al.(2025)Jiang, Ko, Yuan, Shende, and Azenkot}]{chi2025video}
\bibinfo{author}{L.~Jiang}, \bibinfo{author}{W.~Ko}, \bibinfo{author}{S.~Yuan}, \bibinfo{author}{T.~Shende}, \bibinfo{author}{S.~Azenkot},
\newblock \bibinfo{title}{Shifting the focus: Exploring video accessibility strategies and challenges for people with adhd},
\newblock in: \bibinfo{booktitle}{Proceedings of the 2025 CHI Conference on Human Factors in Computing Systems}, CHI '25, \bibinfo{publisher}{Association for Computing Machinery}, \bibinfo{address}{New York, NY, USA}, \bibinfo{year}{2025}. \URLprefix \url{https://doi.org/10.1145/3706598.3713637}. \DOIprefix\doi{10.1145/3706598.3713637}.
\bibitem[{VanLEHN(2011)}]{vanlehn2011its}
\bibinfo{author}{K.~VanLEHN},
\newblock \bibinfo{title}{The relative effectiveness of human tutoring, intelligent tutoring systems, and other tutoring systems},
\newblock \bibinfo{journal}{Educational Psychologist} \bibinfo{volume}{46} (\bibinfo{year}{2011}) \bibinfo{pages}{197--221}. \URLprefix \url{https://doi.org/10.1080/00461520.2011.611369}. \DOIprefix\doi{10.1080/00461520.2011.611369}. \href{http://arxiv.org/abs/https://doi.org/10.1080/00461520.2011.611369}{{\tt arXiv:https://doi.org/10.1080/00461520.2011.611369}}.
\bibitem[{Pardos et~al.(2023)Pardos, Tang, Anastasopoulos, Sheel, and Zhang}]{pardos2023oatutor}
\bibinfo{author}{Z.~A. Pardos}, \bibinfo{author}{M.~Tang}, \bibinfo{author}{I.~Anastasopoulos}, \bibinfo{author}{S.~K. Sheel}, \bibinfo{author}{E.~Zhang},
\newblock \bibinfo{title}{Oatutor: An open-source adaptive tutoring system and curated content library for learning sciences research},
\newblock in: \bibinfo{booktitle}{Proceedings of the 2023 CHI Conference on Human Factors in Computing Systems}, CHI '23, \bibinfo{publisher}{Association for Computing Machinery}, \bibinfo{address}{New York, NY, USA}, \bibinfo{year}{2023}. \URLprefix \url{https://doi.org/10.1145/3544548.3581574}. \DOIprefix\doi{10.1145/3544548.3581574}.
\bibitem[{D'Mello et~al.(2012)D'Mello, Olney, Williams, and Hays}]{dmello2012gazetutor}
\bibinfo{author}{S.~D'Mello}, \bibinfo{author}{A.~Olney}, \bibinfo{author}{C.~Williams}, \bibinfo{author}{P.~Hays},
\newblock \bibinfo{title}{Gaze tutor: A gaze-reactive intelligent tutoring system},
\newblock \bibinfo{journal}{Int. J. Hum.-Comput. Stud.} \bibinfo{volume}{70} (\bibinfo{year}{2012}) \bibinfo{pages}{377–398}. \URLprefix \url{https://doi.org/10.1016/j.ijhcs.2012.01.004}. \DOIprefix\doi{10.1016/j.ijhcs.2012.01.004}.
\bibitem[{Liu et~al.(2024)Liu, Wu, Chen, Li, Kittur, and Myers}]{liu2024selenite}
\bibinfo{author}{M.~X. Liu}, \bibinfo{author}{T.~Wu}, \bibinfo{author}{T.~Chen}, \bibinfo{author}{F.~M. Li}, \bibinfo{author}{A.~Kittur}, \bibinfo{author}{B.~A. Myers},
\newblock \bibinfo{title}{Selenite: Scaffolding online sensemaking with comprehensive overviews elicited from large language models},
\newblock in: \bibinfo{booktitle}{Proceedings of the 2024 CHI Conference on Human Factors in Computing Systems}, CHI '24, \bibinfo{publisher}{Association for Computing Machinery}, \bibinfo{address}{New York, NY, USA}, \bibinfo{year}{2024}. \URLprefix \url{https://doi.org/10.1145/3613904.3642149}. \DOIprefix\doi{10.1145/3613904.3642149}.
\bibitem[{Wen et~al.(2024)Wen, Ping, Wang, Liang, Xu, and Yan}]{wu2024adaptivevoice}
\bibinfo{author}{S.~Wen}, \bibinfo{author}{S.~Ping}, \bibinfo{author}{J.~Wang}, \bibinfo{author}{H.-N. Liang}, \bibinfo{author}{X.~Xu}, \bibinfo{author}{Y.~Yan},
\newblock \bibinfo{title}{Adaptivevoice: Cognitively adaptive voice interface for driving assistance},
\newblock in: \bibinfo{booktitle}{Proceedings of the 2024 CHI Conference on Human Factors in Computing Systems}, CHI '24, \bibinfo{publisher}{Association for Computing Machinery}, \bibinfo{address}{New York, NY, USA}, \bibinfo{year}{2024}. \URLprefix \url{https://doi.org/10.1145/3613904.3642876}. \DOIprefix\doi{10.1145/3613904.3642876}.
\bibitem[{Arapakis and Leiva(2020)}]{arapakis2020mouse}
\bibinfo{author}{I.~Arapakis}, \bibinfo{author}{L.~A. Leiva},
\newblock \bibinfo{title}{Learning efficient representations of mouse movements to predict user attention},
\newblock in: \bibinfo{booktitle}{Proceedings of the 43rd International ACM SIGIR Conference on Research and Development in Information Retrieval}, SIGIR '20, \bibinfo{publisher}{Association for Computing Machinery}, \bibinfo{address}{New York, NY, USA}, \bibinfo{year}{2020}, p. \bibinfo{pages}{1309–1318}. \URLprefix \url{https://doi.org/10.1145/3397271.3401031}. \DOIprefix\doi{10.1145/3397271.3401031}.
\bibitem[{Leiva and Arapakis(2020)}]{leiva2020cursor}
\bibinfo{author}{L.~A. Leiva}, \bibinfo{author}{I.~Arapakis},
\newblock \bibinfo{title}{The attentive cursor dataset},
\newblock \bibinfo{journal}{Frontiers in Human Neuroscience} \bibinfo{volume}{Volume 14 - 2020} (\bibinfo{year}{2020}). \URLprefix \url{https://www.frontiersin.org/journals/human-neuroscience/articles/10.3389/fnhum.2020.565664}. \DOIprefix\doi{10.3389/fnhum.2020.565664}.
\bibitem[{Kasper et~al.(2012)Kasper, Alderson, and Hudec}]{kasper2012wm}
\bibinfo{author}{L.~J. Kasper}, \bibinfo{author}{R.~M. Alderson}, \bibinfo{author}{K.~L. Hudec},
\newblock \bibinfo{title}{Moderators of working memory deficits in children with attention-deficit/hyperactivity disorder (adhd): A meta-analytic review},
\newblock \bibinfo{journal}{Clinical Psychology Review} \bibinfo{volume}{32} (\bibinfo{year}{2012}) \bibinfo{pages}{605--617}. \URLprefix \url{https://www.sciencedirect.com/science/article/pii/S0272735812000979}. \DOIprefix\doi{https://doi.org/10.1016/j.cpr.2012.07.001}.
\bibitem[{Sonne et~al.(2016)Sonne, Marshall, Obel, Thomsen, and Gr\o{}nb\ae{}k}]{sonne2016assistive}
\bibinfo{author}{T.~Sonne}, \bibinfo{author}{P.~Marshall}, \bibinfo{author}{C.~Obel}, \bibinfo{author}{P.~H. Thomsen}, \bibinfo{author}{K.~Gr\o{}nb\ae{}k},
\newblock \bibinfo{title}{An assistive technology design framework for adhd},
\newblock in: \bibinfo{booktitle}{Proceedings of the 28th Australian Conference on Computer-Human Interaction}, OzCHI '16, \bibinfo{publisher}{Association for Computing Machinery}, \bibinfo{address}{New York, NY, USA}, \bibinfo{year}{2016}, p. \bibinfo{pages}{60–70}. \URLprefix \url{https://doi.org/10.1145/3010915.3010925}. \DOIprefix\doi{10.1145/3010915.3010925}.
\bibitem[{Cibrian et~al.(2020)Cibrian, Lakes, Tavakoulnia, Guzman, Schuck, and Hayes}]{cibrian2020wearables}
\bibinfo{author}{F.~L. Cibrian}, \bibinfo{author}{K.~D. Lakes}, \bibinfo{author}{A.~Tavakoulnia}, \bibinfo{author}{K.~Guzman}, \bibinfo{author}{S.~Schuck}, \bibinfo{author}{G.~R. Hayes},
\newblock \bibinfo{title}{Supporting self-regulation of children with adhd using wearables: Tensions and design challenges},
\newblock in: \bibinfo{booktitle}{Proceedings of the 2020 CHI Conference on Human Factors in Computing Systems}, CHI '20, \bibinfo{publisher}{Association for Computing Machinery}, \bibinfo{address}{New York, NY, USA}, \bibinfo{year}{2020}, p. \bibinfo{pages}{1–13}. \URLprefix \url{https://doi.org/10.1145/3313831.3376837}. \DOIprefix\doi{10.1145/3313831.3376837}.
\bibitem[{Sandland(2025)}]{sandland2025rsd}
\bibinfo{author}{B.~Sandland},
\newblock \bibinfo{title}{Neurodivergent experiences of rejection sensitive dysphoria expose the environmental factors too often overlooked},
\newblock \bibinfo{journal}{Neurodiversity} \bibinfo{volume}{3} (\bibinfo{year}{2025}) \bibinfo{pages}{27546330251394516}. \URLprefix \url{https://doi.org/10.1177/27546330251394516}. \DOIprefix\doi{10.1177/27546330251394516}. \href{http://arxiv.org/abs/https://doi.org/10.1177/27546330251394516}{{\tt arXiv:https://doi.org/10.1177/27546330251394516}}.
\bibitem[{Toplak and Tannock(2005)}]{toplak2005time}
\bibinfo{author}{M.~E. Toplak}, \bibinfo{author}{R.~Tannock},
\newblock \bibinfo{title}{Time perception: Modality and duration effects in attention-deficit/hyperactivity disorder},
\newblock \bibinfo{journal}{Journal of Abnormal Child Psychology} \bibinfo{volume}{33} (\bibinfo{year}{2005}) \bibinfo{pages}{639--654}. \DOIprefix\doi{10.1007/s10802-005-6743-6}.
\bibitem[{Eagle et~al.(2024)Eagle, Baltaxe-Admony, and Ringland}]{eagle2024body}
\bibinfo{author}{T.~Eagle}, \bibinfo{author}{L.~B. Baltaxe-Admony}, \bibinfo{author}{K.~E. Ringland},
\newblock \bibinfo{title}{“it was something i naturally found worked and heard about later”: An investigation of body doubling with neurodivergent participants},
\newblock \bibinfo{journal}{ACM Trans. Access. Comput.} \bibinfo{volume}{17} (\bibinfo{year}{2024}). \URLprefix \url{https://doi.org/10.1145/3689648}. \DOIprefix\doi{10.1145/3689648}.
\bibitem[{Sonne et~al.(2016)Sonne, Marshall, Obel, Thomsen, and Gr\o{}nb\ae{}k}]{sonne2016framework}
\bibinfo{author}{T.~Sonne}, \bibinfo{author}{P.~Marshall}, \bibinfo{author}{C.~Obel}, \bibinfo{author}{P.~H. Thomsen}, \bibinfo{author}{K.~Gr\o{}nb\ae{}k},
\newblock \bibinfo{title}{An assistive technology design framework for adhd},
\newblock in: \bibinfo{booktitle}{Proceedings of the 28th Australian Conference on Computer-Human Interaction}, OzCHI '16, \bibinfo{publisher}{Association for Computing Machinery}, \bibinfo{address}{New York, NY, USA}, \bibinfo{year}{2016}, p. \bibinfo{pages}{60–70}. \URLprefix \url{https://doi.org/10.1145/3010915.3010925}. \DOIprefix\doi{10.1145/3010915.3010925}.
\bibitem[{S{\"o}derlund et~al.(2010)S{\"o}derlund, Sikstr{\"o}m, Loftesnes, and Sonuga-Barke}]{soderlund2007noise}
\bibinfo{author}{G.~B.~W. S{\"o}derlund}, \bibinfo{author}{S.~Sikstr{\"o}m}, \bibinfo{author}{J.~M. Loftesnes}, \bibinfo{author}{E.~J.~S. Sonuga-Barke},
\newblock \bibinfo{title}{The effects of background white noise on memory performance in inattentive school children},
\newblock \bibinfo{journal}{Behavioral and Brain Functions} \bibinfo{volume}{6} (\bibinfo{year}{2010}) \bibinfo{pages}{1--10}. \DOIprefix\doi{10.1186/1744-9081-6-55}.
\bibitem[{Tran et~al.(2024)Tran, Lee, and Park}]{tran2024viz}
\bibinfo{author}{T.~Tran}, \bibinfo{author}{H.-N. Lee}, \bibinfo{author}{J.~H. Park},
\newblock \bibinfo{title}{Discovering accessible data visualizations for people with adhd},
\newblock in: \bibinfo{booktitle}{Proceedings of the 2024 CHI Conference on Human Factors in Computing Systems}, CHI '24, \bibinfo{publisher}{Association for Computing Machinery}, \bibinfo{address}{New York, NY, USA}, \bibinfo{year}{2024}. \URLprefix \url{https://doi.org/10.1145/3613904.3642112}. \DOIprefix\doi{10.1145/3613904.3642112}.
\bibitem[{Stefanidi et~al.(2023)Stefanidi, Sch\"{o}ning, Rogers, and Niess}]{stefanidi2023ecosystem}
\bibinfo{author}{E.~Stefanidi}, \bibinfo{author}{J.~Sch\"{o}ning}, \bibinfo{author}{Y.~Rogers}, \bibinfo{author}{J.~Niess},
\newblock \bibinfo{title}{Children with adhd and their care ecosystem: Designing beyond symptoms},
\newblock in: \bibinfo{booktitle}{Proceedings of the 2023 CHI Conference on Human Factors in Computing Systems}, CHI '23, \bibinfo{publisher}{Association for Computing Machinery}, \bibinfo{address}{New York, NY, USA}, \bibinfo{year}{2023}. \URLprefix \url{https://doi.org/10.1145/3544548.3581216}. \DOIprefix\doi{10.1145/3544548.3581216}.
\bibitem[{{European Parliament and Council of the European Union}(2024)}]{euaiact2024}
\bibinfo{author}{{European Parliament and Council of the European Union}}, \bibinfo{title}{{Regulation (EU) 2024/1689 of the European Parliament and of the Council laying down harmonised rules on artificial intelligence (Artificial Intelligence Act)}}, \bibinfo{howpublished}{Official Journal of the European Union L, 2024/1689}, \bibinfo{year}{2024}. \bibinfo{note}{Entered into force August 1, 2024}.
\bibitem[{Kuzilek et~al.(2017)Kuzilek, Hlosta, and Zdrahal}]{kuzilek2017oulad}
\bibinfo{author}{J.~Kuzilek}, \bibinfo{author}{M.~Hlosta}, \bibinfo{author}{Z.~Zdrahal},
\newblock \bibinfo{title}{Open university learning analytics dataset},
\newblock \bibinfo{journal}{Scientific Data} \bibinfo{volume}{4} (\bibinfo{year}{2017}) \bibinfo{pages}{170171}. \DOIprefix\doi{10.1038/sdata.2017.171}.
\bibitem[{Pardos et~al.(2014)Pardos, Baker, San~Pedro, Gowda, and Gowda}]{pardos2014affective}
\bibinfo{author}{Z.~A. Pardos}, \bibinfo{author}{R.~S. Baker}, \bibinfo{author}{M.~San~Pedro}, \bibinfo{author}{S.~M. Gowda}, \bibinfo{author}{S.~M. Gowda},
\newblock \bibinfo{title}{Affective states and state tests: Investigating how affect and engagement during the school year predict end-of-year learning outcomes},
\newblock volume~\bibinfo{volume}{1}, \bibinfo{year}{2014}, pp. \bibinfo{pages}{107--128}. \URLprefix \url{https://learning-analytics.info/index.php/JLA/article/view/3536}. \DOIprefix\doi{10.18608/jla.2014.11.6}.
\bibitem[{Hicks et~al.(2021)Hicks, Stautland, Fasmer, F\o{}rland, Hammer, Halvorsen, Mjeldheim, Oedegaard, Osnes, Gi\ae{}ver~Syrstad, Riegler, and Jakobsen}]{hicks2021hyperaktiv}
\bibinfo{author}{S.~A. Hicks}, \bibinfo{author}{A.~Stautland}, \bibinfo{author}{O.~B. Fasmer}, \bibinfo{author}{W.~F\o{}rland}, \bibinfo{author}{H.~L. Hammer}, \bibinfo{author}{P.~Halvorsen}, \bibinfo{author}{K.~Mjeldheim}, \bibinfo{author}{K.~J. Oedegaard}, \bibinfo{author}{B.~Osnes}, \bibinfo{author}{V.~E. Gi\ae{}ver~Syrstad}, \bibinfo{author}{M.~A. Riegler}, \bibinfo{author}{P.~Jakobsen},
\newblock \bibinfo{title}{Hyperaktiv: An activity dataset from patients with attention-deficit/hyperactivity disorder (adhd)},
\newblock in: \bibinfo{booktitle}{Proceedings of the 12th ACM Multimedia Systems Conference}, MMSys '21, \bibinfo{publisher}{Association for Computing Machinery}, \bibinfo{address}{New York, NY, USA}, \bibinfo{year}{2021}, p. \bibinfo{pages}{314–319}. \URLprefix \url{https://doi.org/10.1145/3458305.3478454}. \DOIprefix\doi{10.1145/3458305.3478454}.
\bibitem[{Kessler et~al.(2005)Kessler, Adler, Ames et~al.}]{kessler2005asrs}
\bibinfo{author}{R.~C. Kessler}, \bibinfo{author}{L.~Adler}, \bibinfo{author}{M.~Ames}, et~al.,
\newblock \bibinfo{title}{The world health organization adult adhd self-report scale (asrs): a short screening scale for use in the general population.},
\newblock \bibinfo{journal}{Psychological Medicine} \bibinfo{volume}{35} (\bibinfo{year}{2005}) \bibinfo{pages}{245--256}. \DOIprefix\doi{10.1017/S0033291704002892}.
\bibitem[{Watson and Watson(2007)}]{watson2007lms}
\bibinfo{author}{W.~R. Watson}, \bibinfo{author}{S.~L. Watson},
\newblock \bibinfo{title}{An argument for clarity: What are learning management systems, what are they not, and what should they become?},
\newblock \bibinfo{journal}{TechTrends} \bibinfo{volume}{51} (\bibinfo{year}{2007}) \bibinfo{pages}{28--34}. \DOIprefix\doi{10.1007/s11528-007-0023-y}.
\bibitem[{Hart and Staveland(1988)}]{hart1988nasatlx}
\bibinfo{author}{S.~G. Hart}, \bibinfo{author}{L.~E. Staveland},
\newblock \bibinfo{title}{Development of nasa-tlx (task load index): Results of empirical and theoretical research} \bibinfo{volume}{52} (\bibinfo{year}{1988}) \bibinfo{pages}{139--183}. \URLprefix \url{https://www.sciencedirect.com/science/article/pii/S0166411508623869}. \DOIprefix\doi{https://doi.org/10.1016/S0166-4115(08)62386-9}.

\end{thebibliography}

\clearpage
\label{TotPages}
\end{document}